\documentclass[conference]{IEEEtran}
\IEEEoverridecommandlockouts

\usepackage{cite}
\usepackage{amsmath,amssymb,amsfonts}
\usepackage{algorithmic}
\usepackage{graphicx}
\usepackage{textcomp}
\usepackage{xcolor}
\usepackage[hyphens]{url}

\def\BibTeX{{\rm B\kern-.05em{\sc i\kern-.025em b}\kern-.08em
    T\kern-.1667em\lower.7ex\hbox{E}\kern-.125emX}}
\begin{document}

\pdfpagewidth=8.5in
\pdfpageheight=11in

\newcommand{\iscasubmissionnumber}{NaN}

\pagenumbering{arabic}

\title{Vorion: A RISC-V GPU with Hardware-Accelerated 3D Gaussian Rendering and Training}
\author{\IEEEauthorblockN{Yipeng Wang, Mengtian Yang, Chieh-pu Lo, and Jaydeep P. Kulkarni}
\IEEEauthorblockA{
\textit{University of Texas at Austin} 
Austin, TX \\
yipeng.wang@utexas.edu, jaydeep@austin.utexas.edu}}

\maketitle
\thispagestyle{plain}
\pagestyle{plain}


\begin{abstract}

3D Gaussian Splatting (3DGS) has recently emerged as a foundational technique for real-time neural rendering, 3D scene generation, volumetric video (4D) capture. However, its rendering and training impose massive computation, making real-time rendering on edge devices and real-time 4D reconstruction on workstations currently infeasible. Given its fixed-function nature and similarity with traditional rasterization, 3DGS presents a strong case for dedicated hardware in the graphics pipeline of next-generation GPUs. This work, Vorion, presents the first GPGPU prototype with hardware-accelerated 3DGS rendering and training. Vorion features scalable architecture, minimal hardware change to traditional rasterizers, z-tiling to increase parallelism, and Gaussian/pixel-centric hybrid dataflow. We prototype the minimal system (8 SIMT cores, 2 Gaussian rasterizer) using TSMC 16nm FinFET technology, which achieves 19 FPS for rendering. The scaled design with 16 rasterizers achieves 38.6 iterations/s for training.

\end{abstract}

\section{Introduction}

The recent emergence of \textit{3D Gaussian Splatting} ~\cite{kerbl2023threedgs} has revolutionized neural scene representation by combining explicit 3D point-based modeling with the efficiency of differentiable Gaussian rendering. 3DGS represents a scene with a set of anisotropic 3D Gaussian primitives, each represented with position, covariance, color, and opacity attributes. The explicit and continuous representation enables real-time photorealistic rendering while maintaining high reconstruction fidelity. Driven by these advantages, 3DGS has rapidly become the foundation for 3D intelligence, including 3D rendering and reconstruction, 3D generation \cite{tang2023dreamgaussian}, Simultaneous Localization and Mapping (SLAM) \cite{keetha2024splatam}, robotic perceptions \cite{kim2015gpmap}, and volumetric video (4D) capture \cite{wu20244d}.

However, deploying 3DGS rendering in real-time and edge scenarios remains challenging. Existing implementations, even on high-end GPUs, struggle to sustain interactive frame rates when rendering large-scale or dynamic scenes~\cite{lumina2025}. On resource-constrained platforms such as the Jetson Orin NX, rendering throughput often falls below 20~FPS, far from the 60–90~FPS requirement of immersive AR/VR applications~\cite{ye2025gaussian}. For example, Mobile Volta GPUs achieve less than 21~FPS on real-world scenes, while the Jetson Orin platform delivers only under 8~FPS on complex datasets such as CoMap~\cite{lumina2025}. 

Meanwhile, training iteratively refines the attributes of millions of Gaussian primitives through backpropagation and poses an even greater computational challenge. On a modern workstation equipped with a single RTX 4090–class GPU, full 3DGS training for a typical 150–300-view scene with 30k iterations generally requires 30–90 minutes, depending on image resolution and Gaussian count. In contrast, server-class hardware such as an H100 or a multi-GPU node can only reduce this to 10–40 minutes, as performance remains constrained by memory bandwidth and host–device communication overhead. Achieving real-time 4D dynamic capture or 3DGS-based SLAM, however, demands continuous reconstruction and radiance-field updates at tens of hertz to keep pace with live sensor streams. This reveals a 100x performance gap between current training pipelines and true real-time capability. 

\begin{figure}[!]       
    \centering
    \includegraphics[width=0.5\textwidth]{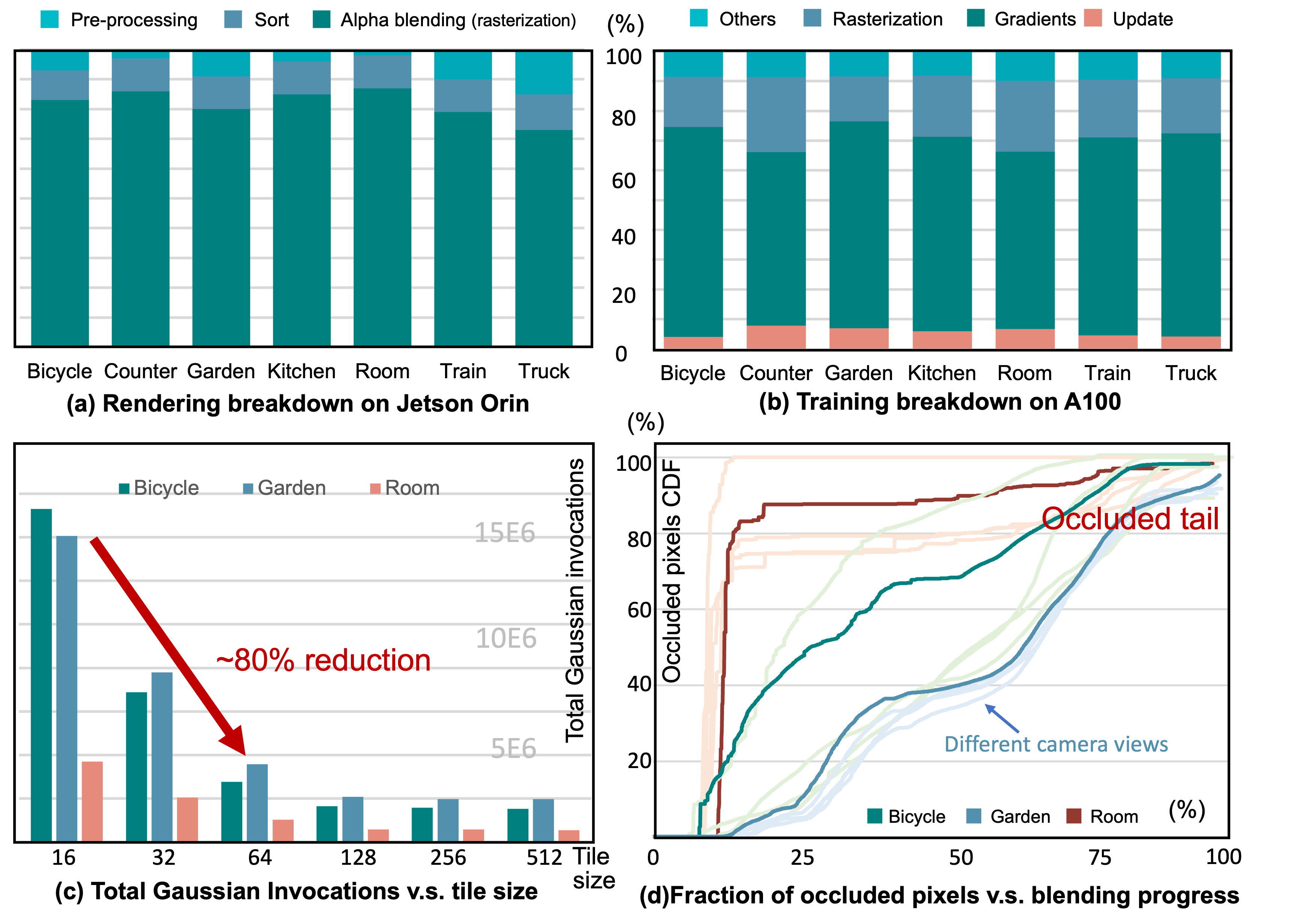}
    \caption{Rendeing and training runtime breakdown on edge and server GPUs; Total Gaussian Invocations v.s. tile size
; Fraction of occluded pixels v.s. blending progress (depth).}
    \label{fig:1}
\end{figure}

\begin{figure*}[t]
  \includegraphics[width=\textwidth]{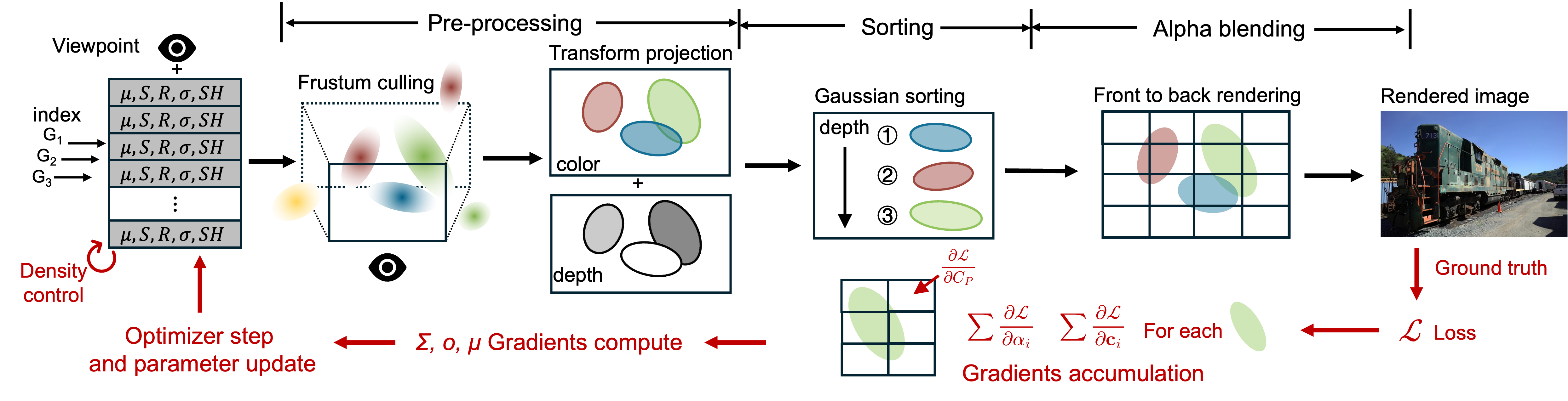}
  \caption{3D Gaussian Splatting rendering and training pipeline.}
  \label{fig:2}
\end{figure*}

Multiple rendering accelerators have been proposed by the architecture and solid-state communities \cite{lee2024gscore,yang2025gsacc, feng20251, li2025gaurast, ye2025gaussian}, and several simulation-based training accelerators have also emerged \cite{gsarch2025,react3d2025}. However, unlike large language models which have massive economic impacts and justify specialized commercial hardware, dedicated accelerators for 3DGS are unlikely to materialize without broader adoption and influence of the algorithm. This creates a paradox: current general-purpose hardware limits the quality and scalability of 3DGS results, yet those limitations hinder the algorithm from achieving the impact needed to motivate custom silicon. Given 3DGS’s fixed-function nature and its strong resemblance to traditional mesh rasterization, there is a compelling opportunity to integrate dedicated support into the graphics pipeline of next-generation GPUs at low cost. We therefore propose augmenting the conventional GPU rasterizer to support both 3DGS rendering and training within the same hardware, reducing silicon cost while maintaining high throughput and scalability. 

In this paper, we first analyze the rendering and training pipelines and identify alpha blending and Gaussian-gradient accumulation as the bottlenecks. We then make two additional observations (Figure. \ref{fig:1}) that motivate using a larger tile size as well as the proposed z-tiling and hybrid dataflow strategies. The detailed Vorion GPU architecture with Gaussian rasterizer and memory subsystem is subsequently presented. Finally, we evaluate performance using a minimal prototype system (8 SIMT cores and 2 Gaussian rasterizers) fabricated in TSMC 16nm FinFET technology, and through post-layout simulation of a scaled configuration featuring 16 rasterizers.

\section{Background}
\subsection{3DGS Rendering}

Rendering in 3D Gaussian Splatting transforms a set of 3D anisotropic Gaussians into screen-space elliptical splats and blends their contributions in visibility order. Figure \ref{fig:2} shows the rendering process which consists of pre-processing, sorting, and alpha-blending.

\textbf{Pre-processing.} In pre-processing, each Gaussian is transformed into camera space, where its depth is extracted for visibility ordering. The 3D covariance, formed from scale parameters and rotation, is projected to a 2D covariance that defines an ellipse on the image plane. Similarly, the mean is projected to pixel coordinates, and a conservative bounding box around the ellipse identifies whether the tiles are touched by the Gaussian. The Gaussian's view-dependent color is evaluated from spherical harmonics, producing a screen-space representation consisting of a 2D mean, a 2D covariance, an opacity, a color, and a depth.

\textbf{Sorting.} For each tile, all overlapping Gaussians are sorted by increasing depth, ensuring that blending proceeds from nearest to farthest. The ordering depends solely on the camera-space depth computed in pre-processing.

\textbf{Alpha-blending.} During alpha-blending, each pixel iterates through the Gaussians assigned to its tile front to back. For a pixel position, the Mahalanobis distance to Gaussian $i$ is computed, and the Gaussian's opacity contribution is
\[
\alpha_{i} = o_i \exp\!\left( -\tfrac{1}{2} \Delta P^\top \Sigma_i^{-1} \Delta P \right),
\]
where $o_i$ is its learned opacity and $\Delta P$ measures the pixel's offset from the projected mean. The pixel's transmittance before encountering Gaussian $i$, and the final pixel color are:
\[
T_{i} = \prod_{j=1}^{i-1} (1 - \alpha_{i}), \ \ \ \ 
C = \sum_{i=1}^n T_{i} \alpha_{i} c_i .
\]

Blending terminates early when transmittance becomes negligible. In this way, pre-processing produces screen-space Gaussians, sorting determines visibility order, and alpha-blending synthesizes the final pixel color.

\subsection{3DGS Training}

\textbf{Forward rasterization.}
Training in 3D Gaussian Splatting begins with the same rasterization procedure used at inference time. Rendering the full image and comparing it to the ground-truth frame produces a pixel-wise loss, whose derivatives with respect to the final pixel colors serve as the entry point for backpropagation.

\textbf{Gradients computation and accumulation.}
Backpropagation distributes these pixel-level gradients to all Gaussians that contributed to the image. Because each pixel depends on a depth-ordered sequence of Gaussians, and each Gaussian influences many pixels, the gradient flow is many-to-many. For Gaussian $i$, let $\mathcal{P}_i$ denote the set of pixels influenced by it. The gradients with respect to its color $\mathbf{c}_i$ and opacity $\alpha_i$ are:
\[
\frac{\partial \mathcal{L}}{\partial \mathbf{c}_i} = 
\sum_{P \in \mathcal{P}_i} 
\left( T_i^P \cdot \alpha_i^P \cdot \frac{\partial \mathcal{L}}{\partial C_P} \right)
\]
\[
\frac{\partial \mathcal{L}}{\partial \alpha_i} = 
\sum_{P \in \mathcal{P}_i} 
\left( 
T_i^P \cdot 
\left( \mathbf{c}_i - \mathbf{C}_{\text{accum}}^{(i,P)} \right)^T 
\cdot 
\frac{\partial \mathcal{L}}{\partial C_P}
-
\frac{T_{\text{final}}^P}{1 - \alpha_i^P} 
\cdot 
\mathbf{C}_{\text{bg}}^T 
\cdot 
\frac{\partial \mathcal{L}}{\partial C_P}
\right)
\]
All other Gaussian gradients, including those for position, scale, rotation, and spherical-harmonic coefficients, are derived by chaining derivatives from these color and opacity gradients. This stage is the primary bottleneck of training. 
Although differentiating every Gaussian–pixel interaction introduces significant computational load, the primary performance bottleneck arises from gradient accumulation: every Gaussian aggregates gradients from potentially thousands of pixels, requiring frequent atomic operations on shared memory buffers. Modern GPUs worsen this effect, as their SM throughput has grown faster than their ROP bandwidth, causing atomic updates to become a serialization point that limits overall training speed.

\textbf{Parameter update.}
After all gradients have been accumulated, the optimizer (typically Adam) updates every Gaussian’s parameters. Because color and opacity gradients propagate through all earlier computations, these updates adjust every aspect of the Gaussian representation, including its geometry, orientation, scale, opacity, and illumination-dependent spherical harmonics.

\textbf{Density control.}
Beyond standard gradient descent, 3DGS employs density control to dynamically adjust the number and placement of Gaussians. Regions that show large residual error or persistent high gradients may trigger cloning or splitting operations, increasing the representational granularity. Conversely, Gaussians that receive consistently small gradients or have near-zero opacity may be removed. This mechanism ensures that model capacity expands as required while avoiding unnecessary growth in areas already well represented.

\section{Observation and Analysis}
\textbf{Alpha blending and gradient accumulation are the dominant bottlenecks.}
Figure \ref{fig:1}(a) and (b) report the rendering and training time breakdowns on the CoMap \cite{yuan2021comap} and
Tanks\;\&\;Temples \cite{knapitsch2017tanks} datasets for the NVIDIA Jetson Orin and A100 platforms, respectively.
Alpha blending consistently accounts for more than 70\% of the total rendering time.
During training, gradient computation and accumulation together exceed 60\% of the
runtime, while rasterization contributes roughly 20\%, again dominated by alpha
blending. In addition, matrix multiplication in Pre-processing and GPU radix sort in Gaussian sorting are scalable and fully optimized in modern GPUs. These observations motivate our proposed Gaussian Rasterizer, 
designed to only accelerate blending as well as gradient computation and accumulation,
while leaving the remaining stages to programmable SIMT cores. Importantly, this
design remains compatible with recent 3DGS ray-tracing \cite{moenne20243d, wu20253dgut} pipelines and other
extensions \cite{huang20242d} in which preprocessing and sorting may differ, but the alpha-blending
stage remains unchanged.

\textbf{Larger tile sizes reduce Gaussian loads.}
Traditional GPU rasterizers typically adopt small tiles (e.g., $16\times16$) to minimize DRAM traffic between pipeline stages. Gaussian splatting, however, does not rely on the multi-stage rasterization pipeline and does not introduce intermediate framebuffer exchanges between stages. Because each Gaussian directly accumulates its contribution into final pixel colors, increasing the tile size does \emph{not} increase DRAM traffic. Instead, it substantially reduces duplicated Gaussian fetches and sorting overhead, since Gaussians that previously overlapped multiple small tiles now fall within a single enlarged tile. As shown in Figure~1 (c), expanding the tile size from $16\times16$ to $64\times64$ reduces per-frame Gaussian invocations by more than $80\%$ across a variety of scenes, with especially pronounced gains on outdoor scenes containing large, spatially extended Gaussian distributions. These reductions directly translate into lower memory bandwidth pressure and improved arithmetic efficiency. Note that our findings differ from those reported in GSCore \cite{lee2024gscore}. The false-positive issue
highlighted in GSCore can be eliminated by a simple intersection 
between each tile’s range and the Gaussian’s bounding box prior to blending.

Motivated by this behavior, we adopt an aggressive $64\times64$ tile size in our architecture. In this work, we use Gaussian-centric execution model \cite{yang2025gsacc} for its memory bandwidth benefits. To further increase parallelism and scalability, we propose z-tiling.

\textbf{Z-tiling.}
In addition to spatial tiling, we also introduce a new form of parallelism based on the depth axis, which we call \emph{z-tiling}. Unlike spatial tiling, which partitions the image plane, z-tiling partitions the \emph{sorted Gaussian list} itself along the depth dimension, enabling parallel or tiled execution while preserving the correct front-to-back compositing semantics. To support this, the depth-sorted set of Gaussians $\{1,\dots,N\}$ is partitioned into $K$ consecutive \emph{z-tiles}, with tile $k$ containing the index range
\[
\mathcal{G}^{(k)}=\{n_{k-1}{+}1,\dots,n_k\}.
\]
Each z-tile independently performs local volumetric compositing,  producing a local color $C_{\mathrm{loc}}^{(k)}$ and an internal residual transmittance $\widehat{T}_{\mathrm{out}}^{(k)}$:
\[
C_{\mathrm{loc}}^{(k)}
=
\sum_{i\in\mathcal{G}^{(k)}}
\widehat{T}^{(k)}_i \, \alpha_i \, c_i,
\qquad
\widehat{T}_{\mathrm{out}}^{(k)}
=
\prod_{i\in\mathcal{G}^{(k)}}
(1-\alpha_i),
\]
where
\[
\widehat{T}^{(k)}_i
=
\prod_{\substack{j\in\mathcal{G}^{(k)} \\ j<i}}
(1-\alpha_j)
\]
is the transmittance accumulated \emph{within} depth tile $k$. To merge the tiles back into a globally consistent front-to-back order, we define an inter-tile transmittance $T_{\mathrm{in}}^{(k)}$, initialized as $T_{\mathrm{in}}^{(1)}=1$ and updated recursively as
\[
T_{\mathrm{in}}^{(k+1)}
=
T_{\mathrm{in}}^{(k)}\, \widehat{T}_{\mathrm{out}}^{(k)}.
\]
The final pixel color is therefore
\[
C
=
\sum_{k=1}^{K}
T_{\mathrm{in}}^{(k)} \, C_{\mathrm{loc}}^{(k)}.
\]
This z-tiling formulation is mathematically equivalent to a single global front-to-back
compositing sweep, but it enables depth-parallel execution and hardware-efficient
scheduling without altering the underlying Alpha blending model. We note, however,
that z-tiling is not applicable during training, as the backward pass must traverse
Gaussians in back-to-front order. This limitation is not restrictive in practice, since
the training procedure already exhibits substantial data parallelism.

\textbf{Many scenes exhibit a strong occluded tail.}
As demonstrated in Figure 1 (d), a small prefix of the depth-sorted Gaussians accounts for nearly all visible contributions. Beyond a certain depth threshold, more than $90\%$ of pixels in many scenes---especially indoor scenes such as \emph{Room}---are fully occluded. Despite this, a Gaussian-centric \cite{yang2025gsacc} execution model still projects and evaluates all back-layer Gaussians, even though most pixels have already saturated their transmittance. This results in substantial wasted work.

Observing this, we propose an optional heterogeneous architecture and hybrid dataflow for edge rendering. For the early z-tiles, where most pixels remain visible, Gaussian-centric processing is ideal: each Gaussian updates all influenced pixels efficiently. Once the system enters the occluded tail, we switch to a pixel-centric mode. In this region, most pixels have negligible remaining transmittance, allowing them to terminate early without evaluating additional Gaussians. This hybrid approach reduces compute and memory pressure on deeply occluded layers, particularly in dense indoor scenes where the occlusion curve saturates rapidly.

Together, with the combination of large spatial tiles, depth-axis z-tiling, and an optional hybrid Gaussian-/pixel-centric dataflow, we present Vorion architecture.

\section{Vorion architecture}
\begin{figure}       
    \centering
    \includegraphics[width=0.49\textwidth]{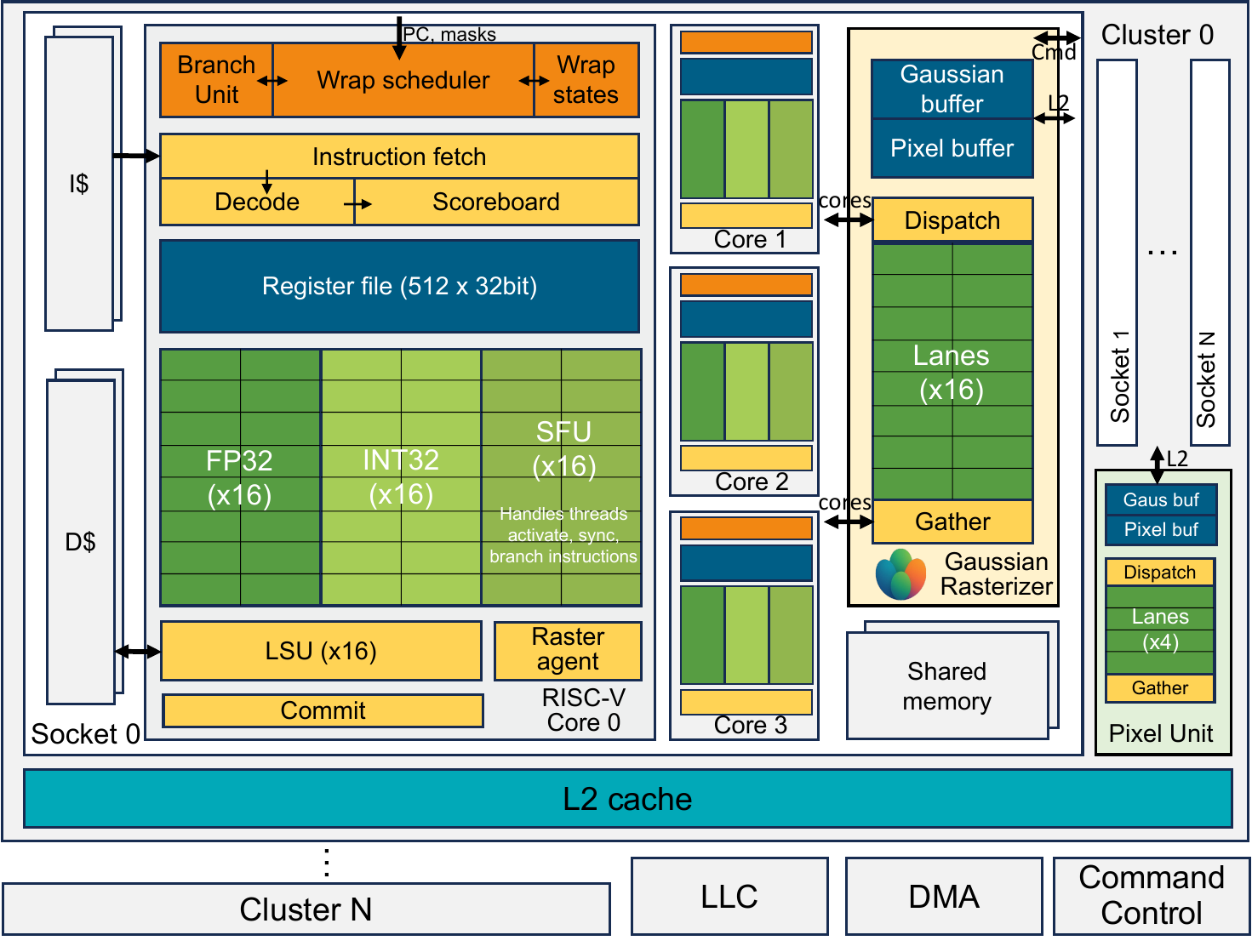}
    \caption{Overall Vorion GPGPU architecture.}
    \vspace{-6pt}
    \label{fig:architecture}
\end{figure}

\subsection{Overall Architecture}
Figure \ref{fig:architecture} presents the overall architecture of our system, built atop the RISC-V GPGPU Vortex platform~\cite{tine2021vortex,tine2023skybox}. 
The design consists of a command controller, a DMA engine, a last-level cache, and multiple compute clusters. 
Each cluster integrates several sockets, a pixel unit, and an associated L2 cache slice. 
Within each socket, there are four RISC-V SIMT cores, a Gaussian rasterizer, and a shared memory block.

Each SIMT core implements a five-stage in-order pipeline and features a warp scheduler capable of launching up to 16 threads sharing a common program counter (PC). 
The register file is single-context, comprising $16\times32$ registers. 
The execution unit provides 16 INT32/FP32 ALUs and 16 SFUs for special arithmetic, branch handling, and thread synchronization.
A lightweight \emph{Raster Agent} coordinates data movement between the register file and the rasterizer buffers through CSR interfaces, managing dispatch and result collection for the Gaussian rasterizer with minimal software overhead.

\subsection{Rendering}
\begin{figure}       
    \centering
    \includegraphics[width=0.49\textwidth]{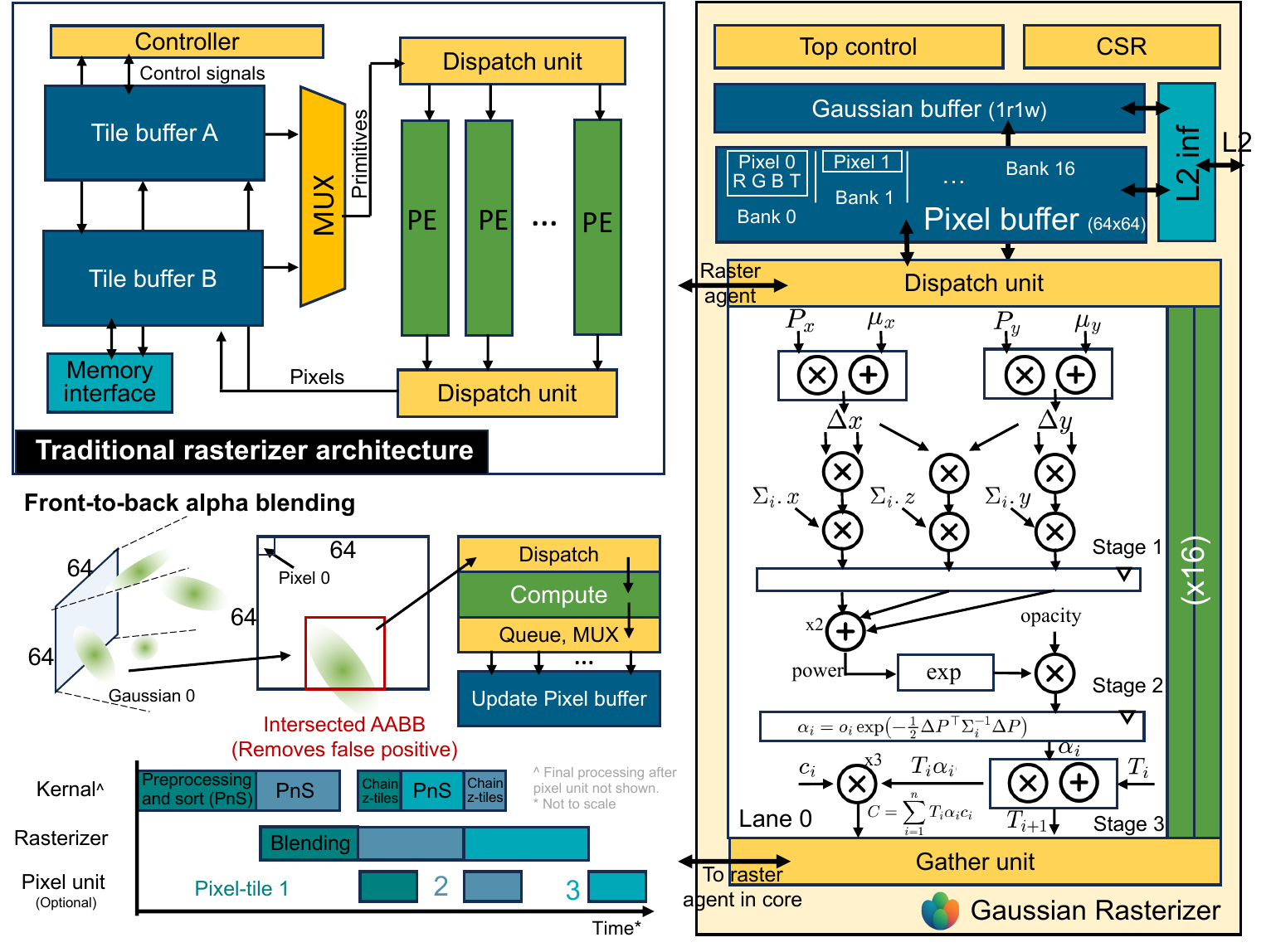}
    \caption{Traditional rasterizer architecture; Gaussian rasterizer architecture in rendering setup; Dataflow chart for rendering.}
    \vspace{-6pt}
    \label{fig:render}
\end{figure}
During the forward pass, programmable kernels handle view transformations, spherical-harmonic color evaluation, and Gaussian sorting on a per-tile basis. 
Depth tiles (z-tiles) are then distributed across the rasterizers, and rendered pixels are written to the frame buffer or directly to DRAM.

\textbf{Gaussian rasterizer architecture.} As shown in Figure \ref{fig:render} Vorion's Gaussian rasterizer intentionally mirrors the structure of conventional triangle rasterizers. 
It features a 1R1W Gaussian buffer, a 1R1W pixel buffer, a set of raster lanes, and dispatch/gather units. 
The command processor configures Gaussian address spaces, initializes pixel buffers, and programs control registers prior to execution.

The rasterizer employs a \emph{Gaussian-centric} dataflow: each Gaussian is brought into the on-chip buffer once per tile, avoiding redundant memory traffic. 
The Gaussian buffer stores the parameters $(\mu,\Sigma,o,c)$, hiding L2 latency during rasterization. 
Before blending, each Gaussian's axis-aligned bounding box (AABB) is intersected with the tile region to eliminate false positives, addressing issues previously noted in GSCore~\cite{lee2024gscore}. 
Surviving pixel tasks are dispatched to raster lanes via a streamlined three-stage pipeline, after which updated pixel values are gathered and written back to the pixel buffer.

\textbf{Memory subsystem.}
Unlike a conventional rasterizer \cite{li2025gaurast}, the Gaussian and pixel buffers in our design do not form
a symmetric ping--pong pair. The Gaussian buffer is a small, low-latency structure used
to hide L2 access delays, whereas the pixel buffer is sized to hold the entire $64\times64$
pixel tile. The pixel buffer is banked into 16 independent slices to match the 16 raster
lanes. Because adjacent pixels tend to be updated in close succession, we use a staggered
banking layout to mitigate conflicts and sustain lane-level parallelism.

The combination of large pixel tiles and a Gaussian-centric execution model eliminates
the need for a dedicated raster cache; the rasterizer connects directly to the L2 slice with
high locality. To support cache-aware scheduling, we embed an ``intersect'' bit in the MSB
of each Gaussian tag, indicating whether it spans multiple tiles and helping guide the
replacement policy.

\textbf{Pixel unit.}
A pixel-centric dataflow may be selected to trade latency for throughput under high
occlusion. When the remaining number of Gaussians becomes small or the accumulated
occlusion ratio exceeds a threshold, the tail of the computation is offloaded to the pixel
unit. Architecturally, this unit mirrors the Gaussian rasterizer but employs four lanes, a
larger Gaussian buffer, and a smaller pixel buffer. Each lane processes one pixel by
iterating over the remaining Gaussians until transmittance collapses. In heavily occluded
indoor scenes such as \emph{Room}, offloading the final quarter of Gaussians to the pixel
unit improves throughput by 82.4\%. 

Together, Kernel pre-processing, Gaussian rasterizer, and Pixel unit forms a 3-stage pipeline. The introduction of the pixel unit and pixel-centric dataflow is also due to intention to support 3D Gaussian Ray tracing \cite{moenne20243d,wu20253dgut} in the future.

\subsection{Training}
\begin{figure}       
    \centering
    \includegraphics[width=0.5\textwidth]{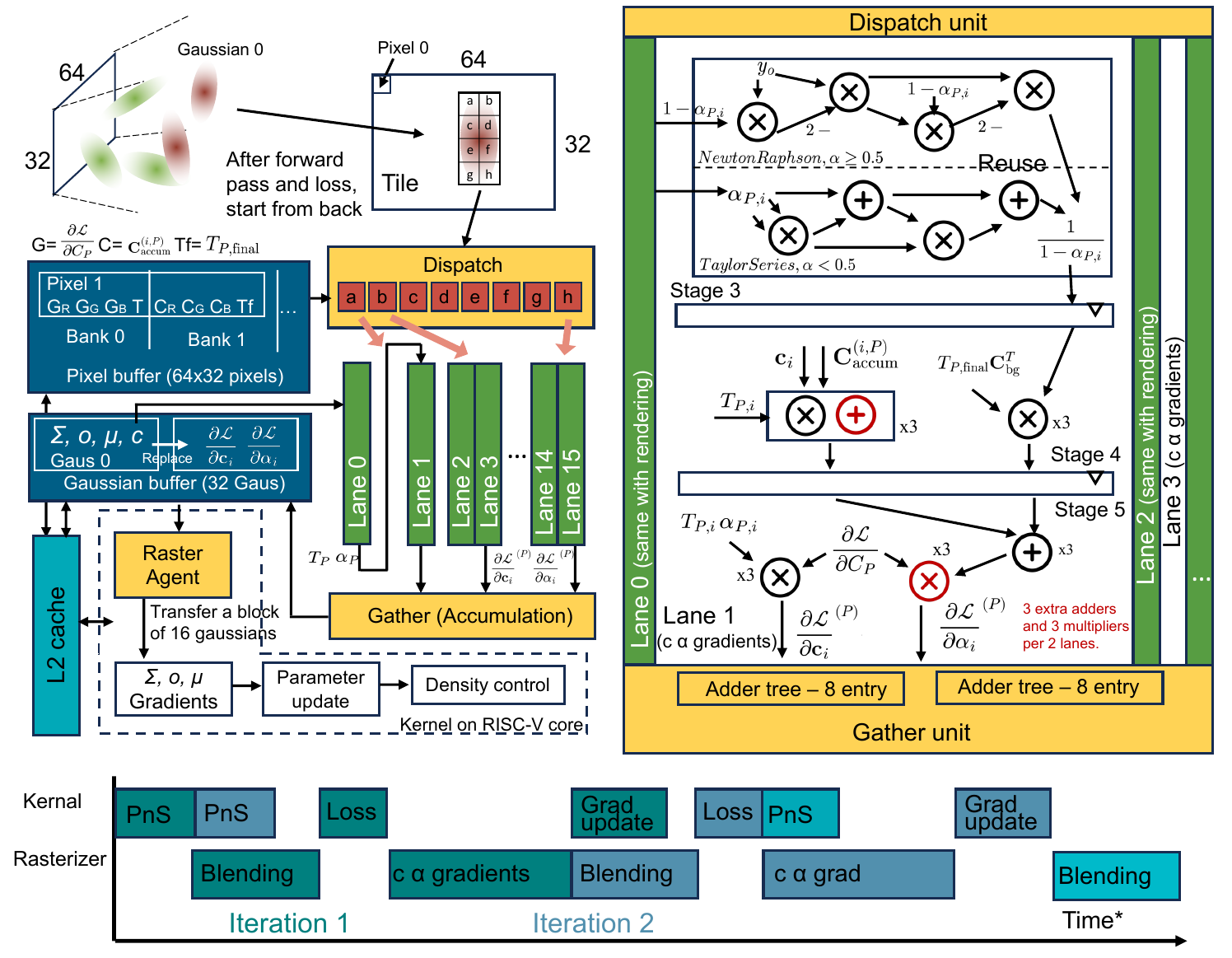}
    \caption{Gaussian rasterizer architecture in training setup; Training dataflow chart.}
    \vspace{-6pt}
    \label{fig:training}
\end{figure}
The Gaussian rasterizer extends naturally to support gradient computation and accumulation for tile-based training, requiring only modest additions to the raster lanes. 
Gradients with respect to Gaussian color $c$ and opacity $\alpha$ account for over 60\% of overall training computation, and the dominant bottleneck is the accumulation of per-pixel gradients back into Gaussian accumulators. 
Consequently, the rasterizer accelerates $c$- and $\alpha$-gradient evaluation directly in hardware, while gradients for $\Sigma$, $o$, and $\mu$ offloaded to programmable kernels.

As illustrated in Figure \ref{fig:training}, kernels first compute the loss and per-pixel gradients. 
Each pixel tile is then assigned to a rasterization engine. 
Per-pixel gradients and the final transmittance $T_{\text{final}}$ are loaded into the pixel buffer, which also tracks the currently accumulated color. 
As gradients need to be stored per pixel, we adopt a $32\times64$ tile shape to keep same buffer dimensions across training and inference.

Gaussians are processed in back-to-front order. 
Even-numbered lanes compute $T_i$ and $\alpha_i$, while chained odd-numbered lanes compute color and opacity gradients. 
A gather unit accumulates these gradients into the per-Gaussian accumulator. 
Every 16 Gaussians, the Raster Agent offloads partial results to the SIMT cores, where kernels compute $\Sigma$, $o$, and $\mu$ gradients. 
For Gaussians intersecting multiple tiles, partial accumulations are computed locally; a cross-tile scan and parameter-update stage (e.g., Adam) is then performed by new kernels.

Because reciprocal hardware is substantially more expensive than multipliers or exponent units, we adopt a hybrid approximation for $(1-\alpha)^{-1}$. 
Empirically, most $\alpha$ values lie below 0.1 or above 0.9. 
For $\alpha<0.5$, we use a fourth-order Taylor approximation; for $\alpha\ge0.5$ (but $<0.99$), we apply two Newton--Raphson iterations, seeded from an 8-entry LUT shared across lanes. 
This approach limits error to within 3\% with no measurable PSNR degradation. 
Supporting these gradient computations adds only three adders and three multipliers per pair of lanes, increasing per-lane area by just 7\%.

\subsection{Software stack}
\begin{figure}[h]       
    \centering
    \includegraphics[width=0.48\textwidth]{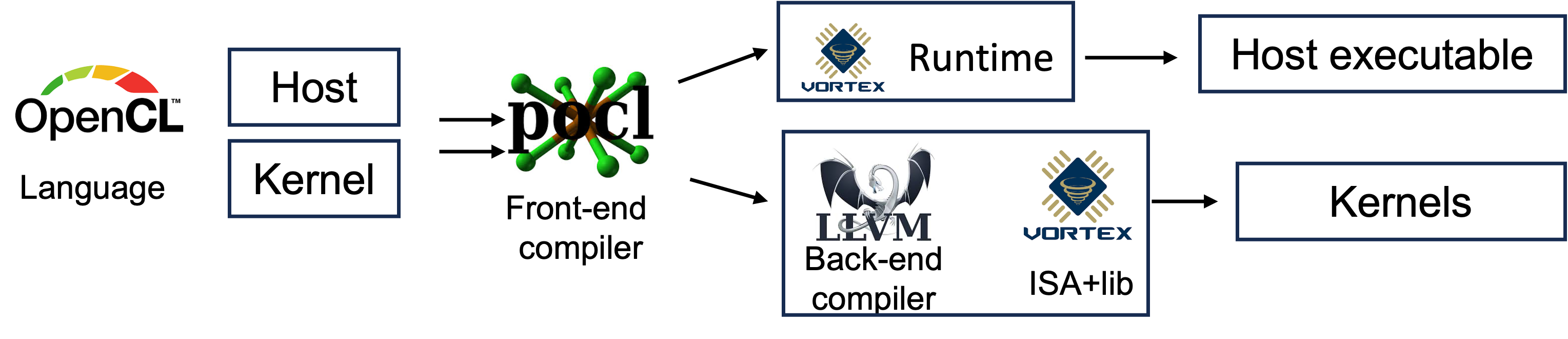}
    \caption{Vorion software stack.}
    \label{fig:soft}
\end{figure}
The 3DGS implementation follows the official CUDA, with the primary 
algorithmic difference being the use of AABB tile intersection. 
All GPU-side Gaussian rasterizer operations are exposed through memory-mapped CSRs, 
allowing kernels to configure rasterizer state, launch raster operations, and retrieve 
results without any specialized driver support.

Figure~\ref{fig:soft} illustrates the complete software toolchain. 
Applications are written in OpenCL, consisting of host code and device kernels. 
Both components are compiled using POCL, which serves as the OpenCL front end and 
partitions the program into host and kernel modules. Kernel code is lowered through 
LLVM’s Vortex back end, which emits binaries targeting the Vorion ISA extensions, 
including our rasterizer intrinsics. The host-side OpenCL runtime links against 
the Vortex runtime library, which provides device discovery, memory management, 
kernel scheduling, and CSR access.

At execution time, the host program issues OpenCL commands through the POCL runtime, 
which forwards kernel binaries and launch parameters to the Vortex driver. 
The kernel modules execute on SIMT cores while interacting with the Gaussian 
rasterizer through CSR operations (e.g., pushing Gaussian descriptors, initiating 
tile rendering, draining pixel buffers). The final host executable thus consists 
of POCL-generated host code linked with Vortex runtime libraries, while the device 
binary contains LLVM-generated kernel code augmented with native rasterizer commands.

\section{Evaluation}
\subsection{Methodology and Experiment Setup}

To evaluate the performance of Vorion for both edge and server senarios. We develope two setups, including a silicon prototype with 8 SIMT cores + 2 rasterizer configuration, and a post layout simulation environment with 64 SIMT core + 16 rasterizer.

\begin{figure}[h]       
    \vspace{-6pt}
    \centering
    \includegraphics[width=0.49\textwidth]{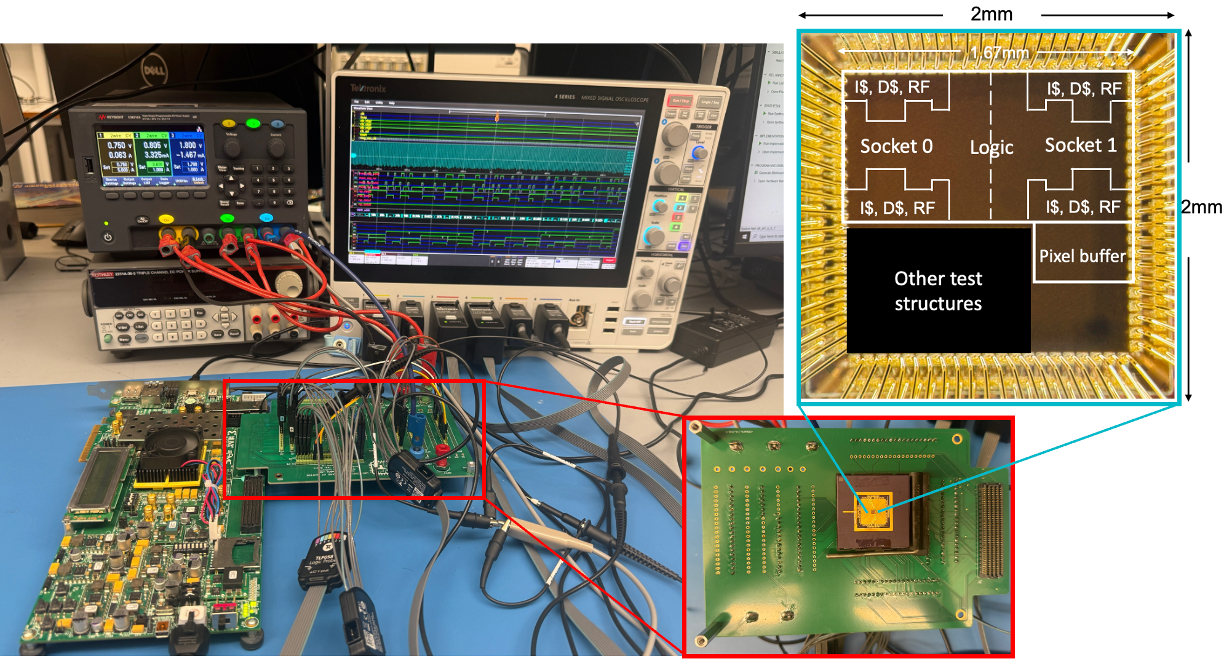}
    \caption{Silicon die shot and prototype setup.}
    \vspace{-6pt}
    \label{fig:proto}
\end{figure}

\textbf{Silicon prototype.}
Our edge-oriented silicon prototype consists of two sockets, each integrating four SIMT
cores and one Gaussian rasterizer, all operating in FP32 precision. The chip is fabricated
in TSMC 16\,nm FinFET technology, occupying a 4\,mm$^2$ die with a 1.6\,mm$^2$ core area, and
is packaged using a cPGA package. System-level functionality---including the command
processor, L2 cache, and DRAM controller---is implemented on an AMD VC707 FPGA. The FPGA
interfaces with the silicon through an FMC connector and provides 12.8\,GB/s of DRAM
bandwidth. 

We evaluate the prototype across supply voltages from 0.57\,V to 1.15\,V and operating
frequencies from 100\,MHz to 530\,MHz. High-frequency clocks are generated on-chip using a
63-stage ring oscillator, which runs at 2.12\,GHz at nominal voltage (0.8\,V) and is subsequently divided
down. Lower-frequency clocks are generated on the FPGA to allow finer-grained control
during characterization.

\textbf{Scaled design.}
The scaled configuration, featuring 64 SIMT cores and 16 rasterizers, uses the same RTL
and technology libraries as the silicon prototype but instantiates more clusters and raster units. The design is
synthesized using Cadence Genus and placed and routed with Cadence Innovus. Post-layout
parasitics are extracted using Cadence Quantus, and timing closure is achieved in Cadence
Tempus at a target frequency of 500\,MHz. Power is estimated with Cadence Voltus using
activity traces collected from representative portions of each benchmark. For system
simulation, we assume an LPDDR4--3200 memory subsystem providing 51.2\,GB/s of bandwidth
modeled using Ramulator~2 \cite{luo2023ramulator}.

\textbf{Performance evaluation.}
\begin{figure}[h]       
    \centering
    \includegraphics[width=0.49\textwidth]{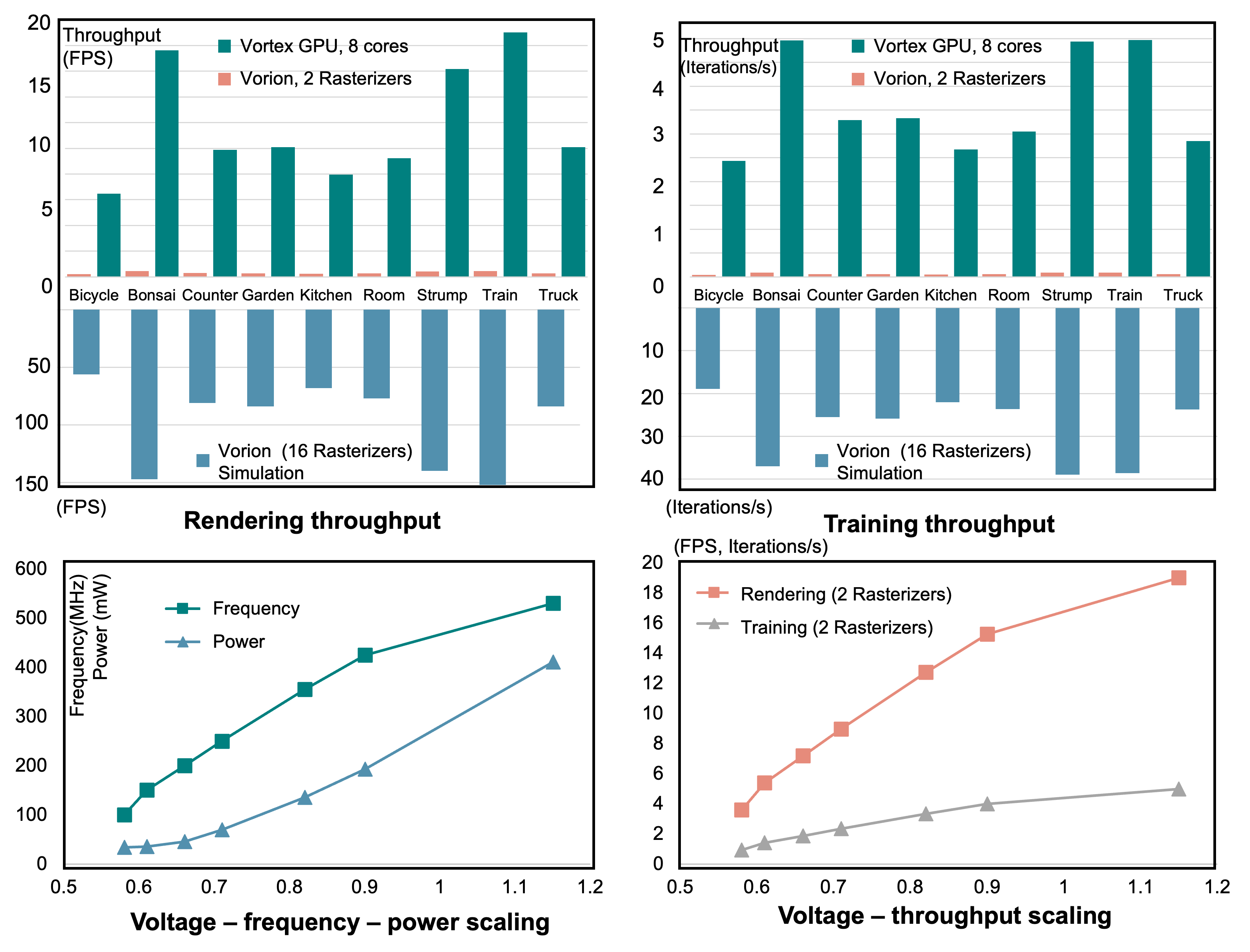}
    \caption{Rendering and training performance evaluation results; Voltage-frequency-power-throughput scaling.}
    \vspace{-9pt}
    \label{fig:perf}
\end{figure}
As Vorion is a general-purpose GPU, we do not compare it to prior fixed-function
accelerators. Instead, we evaluate the impact of the Gaussian rasterizer by comparing
designs with and without rasterization support. As shown in Figure \ref{fig:perf}, across supply voltages from 0.57\,V to 1.15\,V and frequencies from 100\,MHz to 530\,MHz,
the silicon prototype delivers 6.4--19\,FPS for rendering and 2.43--4.97\,iterations/s for
training using two rasterizers. These results correspond to a 30.2--38.4$\times$ speedup in
rendering throughput and a 51--58.4$\times$ speedup in training throughput relative to the
GPU-kernel implementation running on eight SIMT cores without rasterization support.
Figure~\ref{fig:perf} shows that both rendering and training performance scale
quasi-linearly with voltage and frequency, while the measured power remains below 600\,mW
across all operating points.

\begin{figure}[b]       
    \vspace{-6pt}
    \centering
    \includegraphics[width=0.48\textwidth]{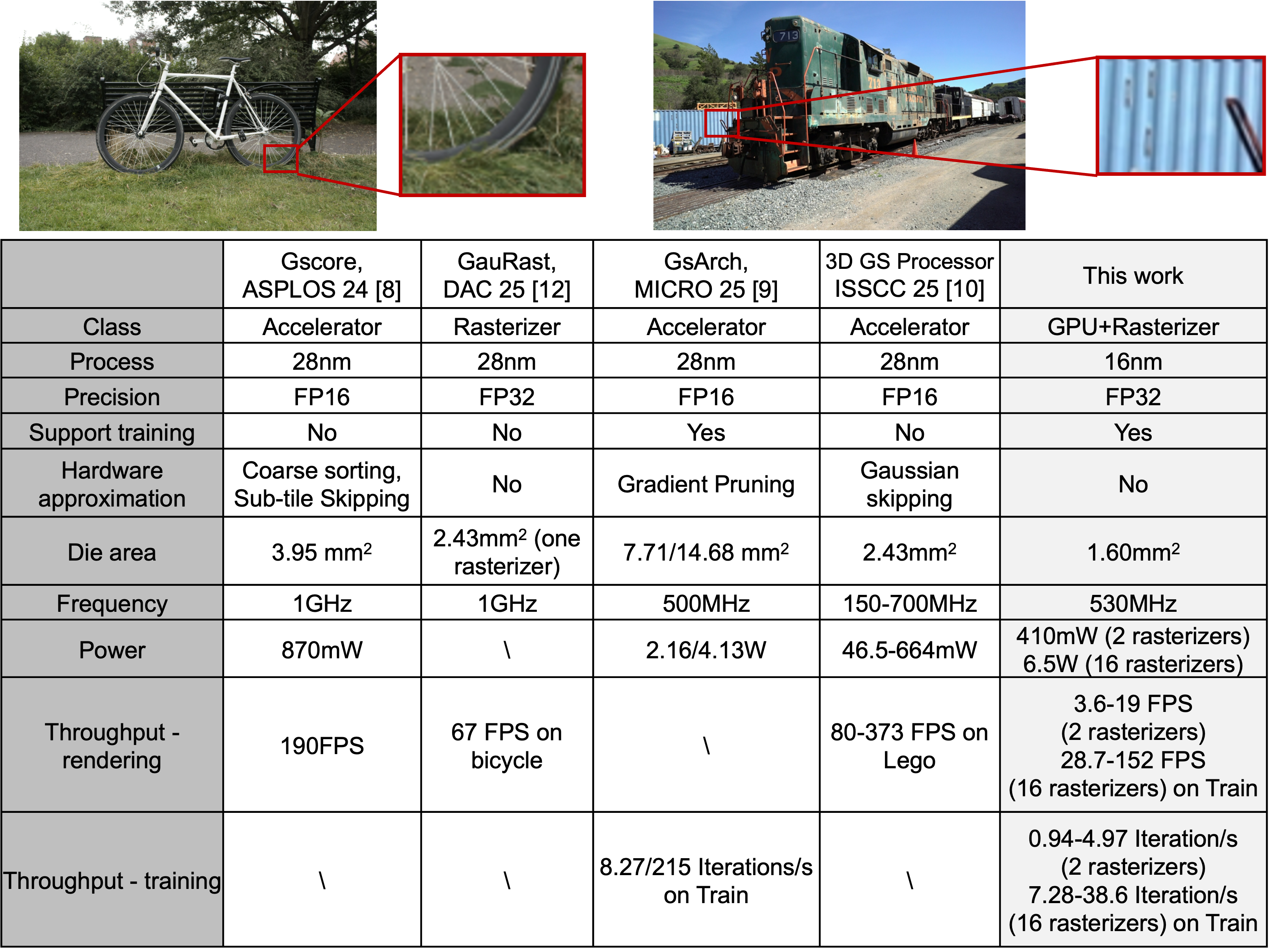}
    \caption{Images rendered by Vorion; Comparison table with prior arts.}
    \vspace{-12pt}
    \label{fig:comp}
\end{figure}

The scaled configuration reaches up to
152\,FPS for rendering and 38.6\,iterations/s for training, demonstrating that our
Gaussian-centric pipeline maintains high efficiency as rasterization resources are
replicated, without limiting by memory bandwidth.

Finally, extrapolating the measured and simulated scaling trends to a configuration with
1{,}024 rasterizers—well within the integration density of modern multi-cluster
accelerators—yields end-to-end training times below 3\,s for a full 3DGS optimization
pass. With future integration of quantization and compression techniques \cite{navaneet2024compgs, papantonakis2024reducing}, which are out of scope of this work, this gap can be
further narrowed, pushing the system toward true real-time operation.

Compared to state-of-the-art 3DGS accelerators~\cite{lee2024gscore, he2025gsarch, feng20251}, as shown in Figure \ref{fig:comp}, the scaled design achieves competitive throughput while preserving full programmability, FP32 precision, and zero accuracy loss. Relative to GPU rasterizer designs~\cite{li2025gaurast}, Vorion achieves 1.7$\times$ normalized area efficiency (FPS/mm$^2$/MHz) per rasterizer.

\section{Conclusion}
This work presented Vorion, a RISC-V–based GPU architecture that introduces the first unified rasterizer for both 3DGS rendering and training. By integrating a Gaussian-centric, large-tile processing, z-tiling, and an optional hybrid dataflow, Vorion directly addresses the computational bottlenecks of alpha blending and gradient accumulation. Our silicon prototype demonstrates 30.4-58.2X speedups over SIMT-only execution while preserving full programmability and FP32 accuracy. Scaled evaluations further show near-linear performance growth with additional rasterization units, achieving up to 152 FPS for rendering and 38.6 iterations/s for training. These results highlight Vorion’s efficiency and scalability, indicating a clear path toward real-time 3DGS for next-generation graphics, robotics, and AR/VR systems.


\bibliographystyle{IEEEtranS}
\bibliography{refs}

@inproceedings{kerbl2023threedgs,
  title     = {3D Gaussian Splatting for Real-Time Radiance Field Rendering},
  author    = {Kerbl, Bernhard and Kopanas, Georgios and Leimk{\"u}hler, Thomas and Drettakis, George},
  booktitle = {ACM SIGGRAPH},
  year      = {2023},
  doi       = {10.1145/3588432.3591542}
}

@inproceedings{react3d2025,
  title     = {REACT3D: Real-time Edge Accelerator for Incremental Training in 3D Gaussian Splatting based SLAM Systems},
  author    = {Wang, Hongyi and Zhu, Zhenhua and Zhao, Tianchen and Xiang, Yunfei and Wang, Zehao and Yu, Jincheng and Yang, Huazhong and Xie, Yuan and Wang, Yu},
  booktitle = {IEEE/ACM International Symposium on Microarchitecture (MICRO)},
  year      = {2025},
  doi       = {10.1145/3725843.3756109}
}

@inproceedings{gsarch2025,
  title     = {GSArch: Breaking Memory Barriers in 3D Gaussian Splatting Training via Architectural Support},
  author    = {He, Houshu and Li, Gang and Liu, Fangxin and Jiang, Li and Liang, Xiaoyao and Song, Zhuoran},
  booktitle = {IEEE International Symposium on High Performance Computer Architecture (HPCA)},
  year      = {2025},
  doi       = {10.1109/HPCA61900.2025.00037}
}

@inproceedings{lumina2025,
  title     = {Lumina: Real-Time Neural Rendering by Exploiting Computational Redundancy},
  author    = {Feng, Yu and Lin, Weikai and Cheng, Yuge and Liu, Zihan and Leng, Jingwen and Guo, Minyi and Chen, Chen and Sun, Shixuan and Zhu, Yuhao},
  booktitle = {International Symposium on Computer Architecture (ISCA)},
  year      = {2025},
  doi       = {10.1145/3695053.3731003}
}

@article{tang2023dreamgaussian,
  title={Dreamgaussian: Generative gaussian splatting for efficient 3d content creation},
  author={Tang, Jiaxiang and Ren, Jiawei and Zhou, Hang and Liu, Ziwei and Zeng, Gang},
  journal={arXiv preprint arXiv:2309.16653},
  year={2023}
}

@inproceedings{keetha2024splatam,
  title={Splatam: Splat track \& map 3d gaussians for dense rgb-d slam},
  author={Keetha, Nikhil and Karhade, Jay and Jatavallabhula, Krishna Murthy and Yang, Gengshan and Scherer, Sebastian and Ramanan, Deva and Luiten, Jonathon},
  booktitle={Proceedings of the IEEE/CVF Conference on Computer Vision and Pattern Recognition},
  pages={21357--21366},
  year={2024}
}

@inproceedings{kim2015gpmap,
  title={GPmap: A unified framework for robotic mapping based on sparse Gaussian processes},
  author={Kim, Soohwan and Kim, Jonghyuk},
  booktitle={Field and Service Robotics: Results of the 9th International Conference},
  pages={319--332},
  year={2015},
  organization={Springer}
}

@inproceedings{wu20244d,
  title={4d gaussian splatting for real-time dynamic scene rendering},
  author={Wu, Guanjun and Yi, Taoran and Fang, Jiemin and Xie, Lingxi and Zhang, Xiaopeng and Wei, Wei and Liu, Wenyu and Tian, Qi and Wang, Xinggang},
  booktitle={Proceedings of the IEEE/CVF conference on computer vision and pattern recognition},
  pages={20310--20320},
  year={2024}
}

@article{yuan2021comap,
  title={Comap: A synthetic dataset for collective multi-agent perception of autonomous driving},
  author={Yuan, Yunshuang and Sester, Monika},
  journal={The International Archives of the Photogrammetry, Remote Sensing and Spatial Information Sciences},
  volume={43},
  pages={255--263},
  year={2021},
  publisher={Copernicus Publications G{\"o}ttingen, Germany}
}

@article{knapitsch2017tanks,
  title={Tanks and temples: Benchmarking large-scale scene reconstruction},
  author={Knapitsch, Arno and Park, Jaesik and Zhou, Qian-Yi and Koltun, Vladlen},
  journal={ACM Transactions on Graphics (ToG)},
  volume={36},
  number={4},
  pages={1--13},
  year={2017},
  publisher={ACM New York, NY, USA}
}

@inproceedings{lee2024gscore,
  title={Gscore: Efficient radiance field rendering via architectural support for 3d gaussian splatting},
  author={Lee, Junseo and Lee, Seokwon and Lee, Jungi and Park, Junyong and Sim, Jaewoong},
  booktitle={Proceedings of the 29th ACM International Conference on Architectural Support for Programming Languages and Operating Systems, Volume 3},
  pages={497--511},
  year={2024}
}

@article{moenne20243d,
  title={3d gaussian ray tracing: Fast tracing of particle scenes},
  author={Moenne-Loccoz, Nicolas and Mirzaei, Ashkan and Perel, Or and de Lutio, Riccardo and Martinez Esturo, Janick and State, Gavriel and Fidler, Sanja and Sharp, Nicholas and Gojcic, Zan},
  journal={ACM Transactions on Graphics (TOG)},
  volume={43},
  number={6},
  pages={1--19},
  year={2024},
  publisher={ACM New York, NY, USA}
}

@inproceedings{wu20253dgut,
  title={3dgut: Enabling distorted cameras and secondary rays in gaussian splatting},
  author={Wu, Qi and Esturo, Janick Martinez and Mirzaei, Ashkan and Moenne-Loccoz, Nicolas and Gojcic, Zan},
  booktitle={Proceedings of the Computer Vision and Pattern Recognition Conference},
  pages={26036--26046},
  year={2025}
}

@inproceedings{huang20242d,
  title={2d gaussian splatting for geometrically accurate radiance fields},
  author={Huang, Binbin and Yu, Zehao and Chen, Anpei and Geiger, Andreas and Gao, Shenghua},
  booktitle={ACM SIGGRAPH 2024 conference papers},
  pages={1--11},
  year={2024}
}

@inproceedings{yang2025gsacc,
  title={GSAcc: Accelerate 3D Gaussian Splatting via Depth Speculation and Gaussian-centric Rasterization},
  author={Yang, Mengtian and Wang, Yipeng and Lo, Chieh-Pu and Zhang, Xiuhao and Oruganti, Sirish and Kulkarni, Jaydeep P},
  booktitle={2025 62nd ACM/IEEE Design Automation Conference (DAC)},
  pages={1--7},
  year={2025},
  organization={IEEE}
}

@inproceedings{tine2021vortex,
  title={Vortex: Extending the risc-v isa for gpgpu and 3d-graphics},
  author={Tine, Blaise and Yalamarthy, Krishna Praveen and Elsabbagh, Fares and Hyesoon, Kim},
  booktitle={MICRO-54: 54th Annual IEEE/ACM International Symposium on Microarchitecture},
  pages={754--766},
  year={2021}
}

@inproceedings{tine2023skybox,
  title={Skybox: Open-source graphic rendering on programmable risc-v gpus},
  author={Tine, Blaise and Saxena, Varun and Srivatsan, Santosh and Simpson, Joshua R and Alzammar, Fadi and Cooper, Liam and Kim, Hyesoon},
  booktitle={Proceedings of the 28th ACM International Conference on Architectural Support for Programming Languages and Operating Systems, Volume 3},
  pages={616--630},
  year={2023}
}

@article{luo2023ramulator,
  title={Ramulator 2.0: A modern, modular, and extensible dram simulator},
  author={Luo, Haocong and Tu{\u{g}}rul, Yahya Can and Bostanc{\i}, F Nisa and Olgun, Ataberk and Ya{\u{g}}l{\i}k{\c{c}}{\i}, A Giray and Mutlu, Onur},
  journal={IEEE Computer Architecture Letters},
  volume={23},
  number={1},
  pages={112--116},
  year={2023},
  publisher={IEEE}
}

@inproceedings{he2025gsarch,
  title={GSArch: Breaking Memory Barriers in 3D Gaussian Splatting Training via Architectural Support},
  author={He, Houshu and Li, Gang and Liu, Fangxin and Jiang, Li and Liang, Xiaoyao and Song, Zhuoran},
  booktitle={2025 IEEE International Symposium on High Performance Computer Architecture (HPCA)},
  pages={366--379},
  year={2025},
  organization={IEEE}
}

@inproceedings{feng20251,
  title={1.78 mJ/Frame 373fps 3D GS Processor Based on Shape-Aware Hybrid Architecture Using Earlier Computation Skipping and Gaussian Cache Scheduler},
  author={Feng, Xiaoyu and Wang, Hedi and Tang, Chen and Wu, Tongda and Yang, Huazhong and Liu, Yongpan},
  booktitle={2025 IEEE International Solid-State Circuits Conference (ISSCC)},
  volume={68},
  pages={1--3},
  year={2025},
  organization={IEEE}
}

@article{li2025gaurast,
  title={GauRast: Enhancing GPU Triangle Rasterizers to Accelerate 3D Gaussian Splatting},
  author={Li, Sixu and Keller, Ben and Lin, Yingyan Celine and Khailany, Brucek},
  journal={arXiv preprint arXiv:2503.16681},
  year={2025}
}

@inproceedings{ye2025gaussian,
  title={Gaussian Blending Unit: An Edge GPU Plug-in for Real-Time Gaussian-Based Rendering in AR/VR},
  author={Ye, Zhifan and Fu, Yonggan and Zhang, Jingqun and Li, Leshu and Zhang, Yongan and Li, Sixu and Wan, Cheng and Wan, Chenxi and Li, Chaojian and Prathipati, Sreemanth and others},
  booktitle={2025 IEEE International Symposium on High Performance Computer Architecture (HPCA)},
  pages={353--365},
  year={2025},
  organization={IEEE}
}

@inproceedings{navaneet2024compgs,
  title={Compgs: Smaller and faster gaussian splatting with vector quantization},
  author={Navaneet, KL and Pourahmadi Meibodi, Kossar and Abbasi Koohpayegani, Soroush and Pirsiavash, Hamed},
  booktitle={European Conference on Computer Vision},
  pages={330--349},
  year={2024},
  organization={Springer}
}

@article{papantonakis2024reducing,
  title={Reducing the memory footprint of 3d gaussian splatting},
  author={Papantonakis, Panagiotis and Kopanas, Georgios and Kerbl, Bernhard and Lanvin, Alexandre and Drettakis, George},
  journal={Proceedings of the ACM on Computer Graphics and Interactive Techniques},
  volume={7},
  number={1},
  pages={1--17},
  year={2024},
  publisher={ACM New York, NY, USA}
}

\end{document}